\documentclass[aps,pre,showpacs,amsmath,amssymb,amsfonts,twocolumn,lengthcheck]{revtex4-1}

\usepackage{graphicx}
\usepackage{subfigure}
\usepackage{verbatim}
\usepackage{dcolumn}
\usepackage{bm}
\usepackage{epsf}
\usepackage{color}
\usepackage[colorlinks=true,citecolor=blue,linkcolor=blue,urlcolor=blue]{hyperref}
\usepackage{hhline}

\newcommand{\bra}[1]{\left\langle #1\right|}
\newcommand{\ket}[1]{\left|#1\right\rangle}

\newcommand{\tr}[1]{\mathrm{tr}\left\{#1\right\}}

\newcommand{\la}{\left\langle}
\newcommand{\ra}{\right\rangle}

\newcommand{\de}[1]{\delta\left(#1\right)}
\newcommand{\td}{\mathrm{d}}

\newcommand{\etal}{\textit{et al. }}
\newcommand{\e}[1]{\exp{\left(#1\right)}}
\newcommand{\lo}[1]{\ln{\left(#1\right)}}
\newcommand{\id}{\mathbb{I}}

\newcommand{\bla}{bla\\bla\\bla\\bla\\bla}

\newcommand{\PRA}{Phys. Rev. A }

\newcommand{\PRE}{Phys. Rev. E }
\newcommand{\PRL}{Phys. Rev. Lett. }

\newcommand{\RMP}{Rev. Mod. Phys. }

\newcommand{\mb}[1]{\mbox{\boldmath$#1$}}
\newcommand{\mc}[1]{\mathcal{#1}}

\newcommand{\mf}[1]{\mathfrak{#1}}

\DeclareMathOperator*{\sumint}{%
\mathchoice%
  {\ooalign{$\displaystyle\sum$\cr\hidewidth$\displaystyle\int$\hidewidth\cr}}
  {\ooalign{\raisebox{.14\height}{\scalebox{.7}{$\textstyle\sum$}}\cr\hidewidth$\textstyle\int$\hidewidth\cr}}
  {\ooalign{\raisebox{.2\height}{\scalebox{.6}{$\scriptstyle\sum$}}\cr$\scriptstyle\int$\cr}}
  {\ooalign{\raisebox{.2\height}{\scalebox{.6}{$\scriptstyle\sum$}}\cr$\scriptstyle\int$\cr}}
}

\begin{document}

\title{Quantum work statistics of charged Dirac particles in time-dependent fields}

\author{Sebastian Deffner}
\author{Avadh Saxena}
\affiliation{Theoretical Division and Center for Nonlinear Studies, Los Alamos National Laboratory, Los Alamos, NM 87545, USA}

\date{\today}

\begin{abstract}
The quantum Jarzynski equality is an important theorem of modern quantum thermodynamics. We show that the Jarzynski equality readily generalizes to relativistic quantum mechanics described by the Dirac equation. After establishing the conceptual framework we solve a pedagogical, yet experimentally relevant, system analytically. As a main result we obtain the exact quantum work distributions for charged particles traveling through a time-dependent vector potential evolving under Schr\"odinger as well as under Dirac dynamics, and for which the Jarzynski equality is verified. Special emphasis is put on the conceptual and technical subtleties arising from relativistic quantum mechanics.
\end{abstract}


\maketitle

\section{Introduction}

The Jarzynski equality \cite{Jarzynski1997} together with subsequent Nonequilibrium Work Theorems, such as the Crooks fluctuation theorem \cite{Crooks1999}, are undoubtedly among the most important breakthroughs in modern Statistical Physics \cite{OrtizdeZarate2011}. In traditional thermodynamics the only processes that are fully describable are infinitely slow -- \emph{equilibrium} -- processes \cite{callen}. For all realistic, finite-time -- \emph{nonequilibrium} -- processes the second law of thermodynamics constitutes an \emph{inequality}, only stating that some portion of the entropy is irreversibly dissipated into the environment.  Jarzynski showed that for isothermal processes the second law of thermodynamics can be formulated as an \emph{equality}, no matter how far from equilibrium the system is driven \cite{Jarzynski1997}, $\la \e{-\beta W}\ra=\e{-\beta \Delta F}$. Here $\beta$ is the inverse temperature of the environment, and $\Delta F$ is the free energy difference, i.e., the work performed during an infinitely slow process.  The angular brackets denote an average over an ensemble of  finite-time realizations of the process characterized by their nonequilibrium work $W$.

The discovery of these so-called fluctuation theorems effectively opened a new field of contemporary research \cite{Bustamante2005a,Jarzynski2015}. For small, but classical systems the Jarzynski equality is a universally valid theorem \cite{Jarzynski2011a}, which has been experimentally verified in a variety of systems \cite{Liphardt2002,Schuler2005,Saira2012,Berut2013}. For quantum systems, however, the situation is more complicated. The major conceptual obstacle is how to generalize the classical notion of thermodynamic work to the quantum domain. In particular, quantum work is not an observable in the usual sense, as there is no hermitian operator, whose eigenvalues are given by the classical work values \cite{tasaki_2000,kurchan_2000,talkner_2007,campisi_2011,Hanggi2015}.

For isolated quantum systems evolving under unitary dynamics the so-called two-time energy measurement approach has proven to be practical and powerful. In this paradigm, quantum work is determined by projective energy measurements at the beginning and the end of a process induced by an externally controlled Hamiltonian. Although this approach has been verified experimentally \cite{huber_2008,Dorner2013,Mazzola2013a,Batalhao2014,An2014b} and has led to the development of thermodynamic quantum devices \cite{abah_2012,rossnagel_2013,Zhang2014a}, the paradigm cannot be considered entirely satisfactory as it relies on a rather invasive procedure -- projective measurements -- and is restricted to isolated systems.

Thus, modern quantum thermodynamics has been attempting to overcome these restrictions: On the one hand, researchers have generalized the two-time energy measurement approach to less invasive procedures such as generalized measurements \cite{kafri_2012,Venkatesh2013,Watanabe2014a,Allahverdyan2014a,Roncaglia2014,Manzano2015}, or to quantum systems that are less ``isolated'' such as in $\mc{PT}$-symmetric quantum mechanics \cite{Deffner2015a}. On the other hand, various notions of quantum work and entropy production for general, open quantum systems have been proposed \cite{Subas2012,Horowitz2012a,deffner_2013,Campisi2013a,Leggio2013,Leggio2013b}, which however all lack the desired universality of notions from traditional thermodynamics.

Nevertheless, due to its simplicity and practicality for isolated quantum systems a great deal of research has been dedicated to a careful study of the quantum work statistics from two-time energy measurements. For instance, the quantum work distribution has been computed for time-dependent oscillators \cite{deffner_2008,talkner_2008,deffner_2010}, a particle in a time-dependent box \cite{Quan2012}, quantum Ising chains \cite{Silva2008,Smacchia2013,Marino2014,Fusco2014a}, the Landau-Zener model \cite{Mascarenhas2014}, noninteracting bosons and fermions \cite{Gong2014}, diatomic molecules \cite{Leonard2014a}, etc.

However, to the best of our knowledge all prior work has focused on non-relativistic quantum systems, while a generalization of the Jarzynski equality to relativistic energies has only been proposed for classical systems \cite{Fingerle2007}. The present paper aims at closing this gap and reports the generalization of the quantum Jarzynski equality to particles evolving under the time-dependent Dirac equation. We will see that the validity of the Jarzynski equality together with the two-time energy measurement approach follows directly from the unitarity of Dirac dynamics -- the only essential requirement \cite{kafri_2012}. Therefore, after briefly establishing the conceptual building blocks, we will focus on a pedagogical and illustrative case study, namely charged spin-$1/2$ particles traveling through a time-dependent vector potential. 

The purpose of the present study is two-fold: We will show that the quantum Jarzynski equality naturally holds for dynamics described by the Dirac equation. The main part of the following discussion, however, will provide a ``recipe'' of how to compute the relativistic quantum work density. Our analysis will put emphasis on the technical and conceptual subtleties arising from Dirac's equation, and we will compare our relativistic results with the analogous Schr\"odinger dynamics.

\section{Relativistic quantum work}

We begin by briefly reviewing the paradigm of the two-time energy measurement approach, and establish notions and notations. Consider an isolated quantum system with time-dependent Schr\"odinger equation
\begin{equation}
\label{eq01}
i\hbar\,|\dot{\psi}_t\rangle=H_t\,\ket{\psi_t}\,,
\end{equation}
where the dot denotes a derivative with respect to time. We are interested in describing thermodynamic processes that are induced by varying an external control parameter $\lambda_t$ during time $\tau$, with $H_t=H(\lambda_t)$. Within the paradigm of two-time energy measurements quantum work is determined by the following, experimentally motivated protocol: After preparation of the initial state $\rho_0$ a projective energy measurement is performed; then the system is allowed to evolve under the time-dependent Schr\"odinger equation \eqref{eq01}, before a second projective energy measurement is performed at $t=\tau$. Thus, for a single realization of this protocol the quantum work is given by
\begin{equation}
\label{eq02}
W_{n_0\rightarrow n_\tau}=\epsilon(n_\tau,\lambda_{\tau})-\epsilon(n_0,\lambda_0)\,,
\end{equation}
where $\ket{n_0}$ is the initial eigenstate with eigenenergy $\epsilon(n_0,\lambda_0)$ and $\ket{n_\tau} $ with $\epsilon(n_\tau,\lambda_\tau)$ describes the final eigenstate.

The quantum work density is then given by an average over an ensemble of realizations, $\mc{P}(W)=\la \de{W-W_{n_0\rightarrow n_\tau}}\ra$, which can be rewritten as \cite{kafri_2012,deffner_2013}
\begin{equation}
\label{eq03}
\mathcal{P}(W)=\sumint_{n_0,n_\tau} \de{W-W_{n_0\rightarrow n_\tau}}\,\mf{p}\left(n_0\rightarrow n_\tau\right).
\end{equation}
In the latter equation the symbol $\sumint$ denotes a sum over the discrete part of the eigenvalues spectrum and an integral over the continuous part. 

To compute $\mc{P}(W)$ \eqref{eq03} explicitly, one has to determine the transition probabilities  $\mf{p}\left(n_0\rightarrow n_\tau\right)$ first. These can be written as \cite{kafri_2012,Deffner2015a},
\begin{equation}
\label{eq04}
\mf{p}\left(n_0\rightarrow n_\tau\right)=\tr{\Pi_{n_\tau}\, U_{\tau}\, \Pi_{n_0}\,\rho_0\,\Pi_{n_0}\, U_{\tau}^\dagger}\,,
\end{equation}
where $\rho_0$ is the initial density operator of the system and $U_{\tau}$ is the unitary time evolution operator, $U_{\tau}=\mc{T}_> \e{-i/\hbar\,\int_0^{\tau}\td t\,H_t}$. Finally, $\Pi_n$ denotes the projector into the space spanned by the $n$th eigenstate, which becomes for non-degenerate spectra $\Pi_n=\ket{n}\bra{n}$.

It is then a simple exercise to show that from the definition of $\mc{P}(W)$ \eqref{eq03} and for an initial Gibbs state, $\rho_0=\e{-\beta H_0}/Z_0$, we have the quantum Jarzynski equality \cite{kurchan_2000,tasaki_2000,talkner_2007,campisi_2011},
\begin{equation}
\label{eq05}
\la \e{-\beta W}\ra=\e{-\beta \Delta F}\,,
\end{equation}
where $\Delta F=F_\tau-F_0$ and $F_t=-(1/\beta)\,\lo{Z_t}$. 

It is worth emphasizing that the validity of the quantum Jarzynski equality is not restricted to Schr\"odinger dynamics. Rather, it has been shown that Eq.~\eqref{eq05} holds for all quantum systems, whose dynamics is at least unital \cite{kafri_2012,Albash2013,Rastegin2013a,Rastegin2014,Deffner2015a}. Unital dynamics preserves the identity and can be written as a superposition of unitary quantum maps \cite{Nielsen2010}. 

Therefore, to check whether the quantum Jarzynski equality holds for Dirac dynamics, one only has to verify that the corresponding evolution equation describes unital dynamics.

\paragraph*{Relativistic quantum mechanics: Dirac equation}

The Dirac equation is a relativistic wave equation, which describes massive spin-$1/2$ particles, such as electrons and quarks. In its original formulation  for free particles the Dirac equation reads \cite{Dirac1928}
\begin{equation}
\label{eq06}
i\hbar\, \dot{\Psi}(\mb{p},t)=\left(c\, \mb{\alpha}\cdot\mb{p}+\alpha_0\,mc^2\right)\,\Psi(\mb{p},t)\,.
\end{equation}
Here, $\Psi(\mb{p},t)$ is the wave function of an electron with rest mass $m$ and momentum $\mb{p}=(p_1,p_2,p_3)$, and $c$ is the speed of light. In covariant form the matrices $\mb{\alpha}=(\alpha_1,\alpha_2,\alpha_3)$ and $\alpha_0$ can be expressed as \cite{Peskin1995},
\begin{equation}
\label{eq07}
\alpha^0=\gamma^0\quad\mathrm{and}\quad\gamma^0\, \alpha^k=\gamma^k\,.
\end{equation}
The $\gamma$-matrices are commonly expressed in terms of $2\times 2$ sub-matrices with the Pauli-matrices $\sigma_x, \sigma_y, \sigma_z$ and the identity $\id_2$ as,
\begin{equation}
\label{eq08}
\begin{split}
\gamma^0=\begin{pmatrix}\id_2&0\\0&-\id_2 \end{pmatrix}\quad&\gamma^1=\begin{pmatrix}0&\sigma_x\\-\sigma_x&0 \end{pmatrix}\\
\gamma^2=\begin{pmatrix}0&\sigma_y\\-\sigma_y&0 \end{pmatrix}\quad&\gamma^3=\begin{pmatrix}0&\sigma_z\\-\sigma_z&0 \end{pmatrix}\,.
\end{split}
\end{equation}
It is then easy to see that the right side of Eq.~\eqref{eq06}, i.e., the operator $c\, \mb{\alpha}\cdot\mb{p}+\alpha_0\,mc^2$, is hermitian, and consequently $\Psi(\mb{p},t)$ evolves under unitary dynamics.

Therefore, the quantum Jarzynski equality \eqref{eq05} also holds for particles evolving under Dirac dynamics \eqref{eq06}. However, we expect the work density function \eqref{eq03} to be dramatically different: In contrast to the Schr\"odinger equation \eqref{eq01} the Dirac wave function $\Psi(\mb{p},t)$ is a bispinor, which can be interpreted as a superposition of a spin-up electron, a spin-down electron, a spin-up positron, and a spin-down positron \cite{Peskin1995}. In addition, the momentum of Dirac particles is confined by the light cone, whereas Schr\"odinger particles can travel with arbitrary velocities.

In the remainder of this study we will analyze the consequences of relativistic effects on the quantum work density for a simple, yet elucidating example. 

\section{Charged particles in a time-dependent vector field}

For the sake of simplicity we now restrict ourselves to a one-dimensional system in $x$-direction. In this case the 4-component Dirac spinor can be separated into two identical 2-component bispinors, which evolve under \cite{Fillion2012},
\begin{equation}
\label{eq09}
i\hbar\, \dot{\Psi}(p,t)=\left(c p\,\sigma_x+ m c^2\, \sigma_z\right)\,\Psi(p,t)\,,
\end{equation}
with $p_x\equiv p$. We further assume that the system is driven by a time-dependent, but spatially homogeneous vector potential $A_t$. For oscillating $A_t$ this situation has been recently solved analytically \cite{Fillion2012}. Moreover, Eq.~\eqref{eq09} describes particle-antiparticle production in counterpropagating laser light, which has been proposed to be observable in an experiment \cite{Fillion2012}. 

Note that the Dirac equation \eqref{eq09} as any electromagentic theory is gauge invariant. Here, ``gauge invariance'' means that a whole class of scalar and vector potentials, related by so-called gauge transformations, describes the same physical situation. In particular, the dynamics of the electromagnetic fields and the dynamics of a charged system in an electromagnetic background do not depend on the choice of the gauge. In the present context this means that the energy eigenvalues \eqref{eq11} and \eqref{eq19} do depend on the gauge, whereas the Jarzysnki equality is gauge invariant \cite{campisi_2011}.

\subsection{Schr\"odinger dynamics}

To build intuition and as a point of reference we treat the non-relativistic problem, first.  In this case the dynamics is described by the time-dependent Schr\"odinger equation, which reads in momentum representation
\begin{equation}
\label{eq10}
i\hbar\, \dot{\psi}(p,t)=\frac{1}{2 m}\,\left(p+A_t/c\right)^2\,\psi(p,t)\,,
\end{equation}
where $A_t$ is the vector potential, and we work in units for which the elementary charge is set to one, $e=1$. 

Accordingly, the instantaneous eigenenergies are,
\begin{equation}
\label{eq11}
\epsilon_S(\pi_t, A_t)=\frac{1}{2m}\,\left(\pi_t+A_t/c\right)^2
\end{equation}
with the corresponding eigenstates,
\begin{equation}
\label{eq12}
\phi(p,\pi_t)=\de{p-(\pi_t+A_t/c)}\,.
\end{equation} 
Here and in the following $\pi_t$ denotes the quantum number, which is in the present case reduces to the eigenmomentum. Note that the eigenstates \eqref{eq12} form an orthonormal basis, since
\begin{equation}
\label{eq13}
\int\td p\,\phi(p,\pi_1) \phi(p,\pi_2)=\de{\pi_1-\pi_2}\,.
\end{equation}
In the present case the time-dependent Schr\"odinger equation \eqref{eq10} reduces to an ordinary differential equation of first order. Thus, a solution of Eq.~\eqref{eq10} can be written as
\begin{equation}
\label{eq14}
\psi(p,t)=\e{-\frac{i}{\hbar}\,\int_0^t \td t'\,\frac{\left(p+A_{t'}/c\right)^2}{2m}}\,\psi(p,0)\,,
\end{equation}
which follows from inspection.

Notice that in the case of Schr\"odinger dynamics the effect of $A_t$ manifests itself exclusively as a time-dependent phase \eqref{eq14}. Shortly, we will see that for the corresponding Dirac equation the situation is much more involved.

\begin{figure}
\includegraphics[width=.48\textwidth]{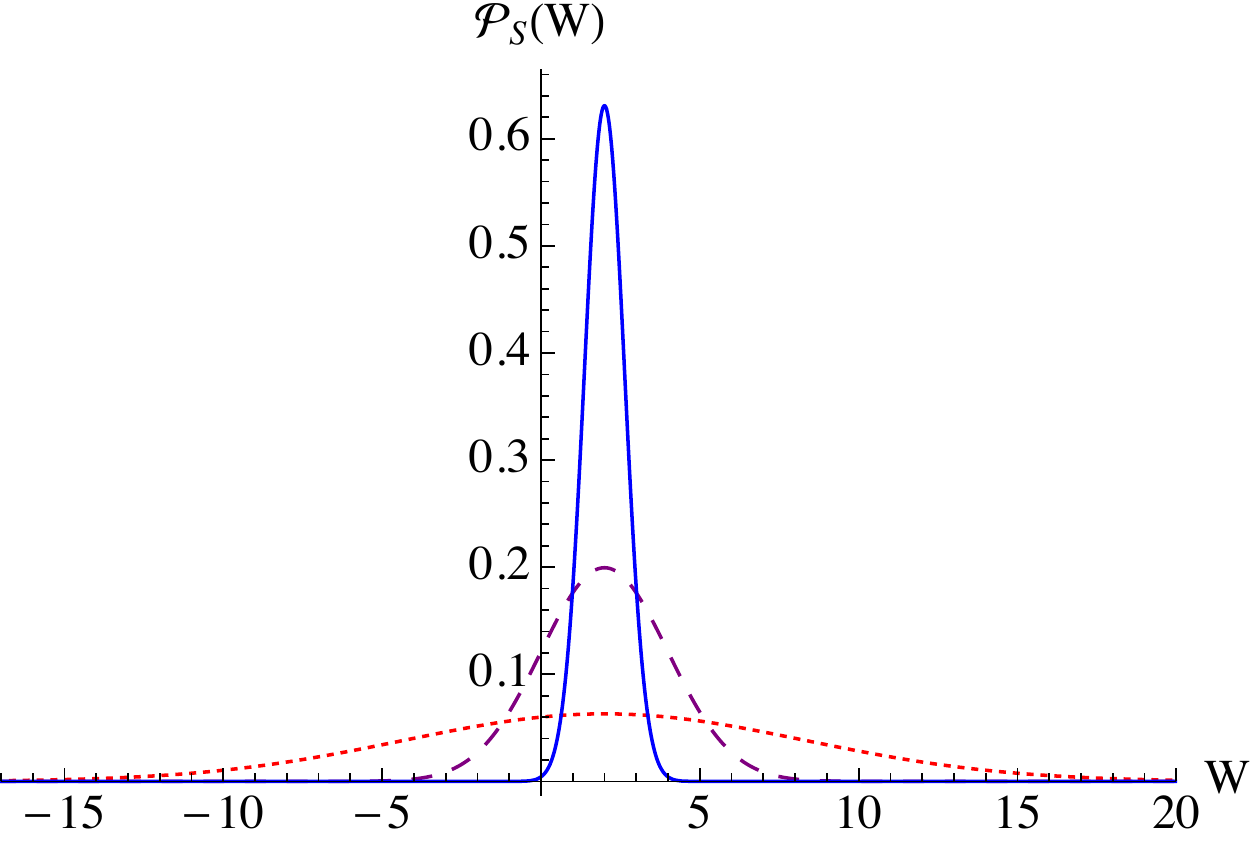}
\caption{\label{fig:schrodinger} (color online) Quantum work density, $\mc{P}_S(W)$, for charged Schr\"odinger particles \eqref{eq16} with $A_0=0$, $A_\tau=1$, $m=1$, $c=1$, and $\beta=10$ (blue, solid line), $\beta=1$ (purple, dashed line), and $\beta=0.1$ (red, dotted line).}
\end{figure}

The instantaneous eigenenergies \eqref{eq11} together with the eigenstates \eqref{eq12} and the time-dependent solution \eqref{eq14} are all ingredients necessary to compute the quantum work density \eqref{eq03}. In particular, the transition probabilities \eqref{eq04} become,
\begin{equation}
\label{eq15}
\mf{p}_S\left(\pi_0\rightarrow\pi_\tau\right)=\left|\int\td p\,\phi(p,\pi_\tau)\,\psi(p,t)\right|^2 \mf{p}_0^S(\pi_0)\,,
\end{equation}
with the initial state
\begin{equation}
\label{eq15a}
\mf{p}_0^S(\pi_0)= \e{-\beta\, \epsilon_S(\pi_0,A_0)}/Z^S_0\,,
\end{equation}
and partition function $Z^S_0=\int\td\pi_0\,\e{-\beta\, \epsilon_S(\pi_0,A_0)}$ and $\psi(p,0)\equiv \phi(p,\pi_0)$. Substituting Eqs.~\eqref{eq11} and \eqref{eq15} into Eq.~\eqref{eq03} we finally obtain after a few lines of simple algebra,
\begin{equation}
\label{eq16}
\begin{split}
&\mc{P}_S(W)=\\
&\frac{\sqrt{\beta\,m c^2}}{ \sqrt{2 \pi}\, \left| A_0- 2 A_\tau\right|}
  \e{-\frac{\beta \left[2  m c^2\, W-(A_0-2 A_\tau)^2\right]^2}{8 m c^2\,(A_0 - 2 A_\tau)^2}}\,.
 \end{split}
\end{equation}
Equation~\eqref{eq16} constitutes our first main result. The quantum work distribution for charged Schr\"odinger particles traveling trough a time-dependent vector potential, $A_t$, is a Gaussian, which is fully determined by the initial and final value of $A_t$. In particular, $\mc{P}_S(W)$ is \emph{independent} of the specific protocol, as $A_t$ merely induces a time-dependent phase \eqref{eq14}. As a point of reference and for comparison with the Dirac case in the following subsection, we plot Eq.~\eqref{eq16} in Fig.~\ref{fig:schrodinger} for low, intermediate, and high temperatures.

\subsection{Dirac dynamics}

In complete analogy to the preceding Schr\"odinger case we now compute the quantum work density for charged particles evolving under the time-dependent Dirac equation,
\begin{equation}
\label{eq17}
i\hbar\, \dot{\Psi}(p,t)=\left[\left(c p+A_t\right)\,\sigma_x+ m c^2\, \sigma_z\right]\,\Psi(p,t)\,.
\end{equation}
Equation~\eqref{eq17} can be separated into two evolution equations for the components of the bispinor, $\Psi(p,t)=\left(\Psi_1(p,t),\Psi_2(p,t)\right)$, and we have
\begin{equation}
\label{eq18}
\begin{split}
\hbar^2\,\ddot{\Psi}_1(p,t)&=-\left[(c p + A_t)^2 + \left(m c^2\right)^2 +i\hbar \,\dot{A_t}\right]\,\Psi_1(p,t)\,,\\
\Psi_2(p,t)&=\left[i\hbar\,  \dot{\Psi}_1(p,t)- (c p + A_t) \Psi_1(p,t)\right]/mc^2\,.
\end{split}
\end{equation}
In contrast to the previous case \eqref{eq10} the solution of the time-dependent Dirac equation \eqref{eq17} is determined by an ordinary differential equation of second order \eqref{eq18}. Therefore, to find analytical solutions of Eq.~\eqref{eq17} we have to resort to particular parameterizations of $A_t$. For oscillating protocols Eq.~\eqref{eq17} has been solved in Ref.~\cite{Fillion2012}, and we will see two further examples in the following.

Before we turn to specific parameterizations, however, we note the instantaneous (positive) eigenenergies of Eq.~\eqref{eq17},
\begin{equation}
\label{eq19}
\epsilon_D(\pi_t, A_t)=\sqrt{\left(c \pi_t+A_t\right)^2+\left(m c^2\right)^2}
\end{equation}
and the corresponding, orthonormal eigenstates,
\begin{equation}
\label{eq20}
\Phi(p,\pi_t)=\frac{\de{p-(\pi_t+A_t/c)}}{ \sqrt{1 + \left(\sqrt{\Pi_t^2 + 1} - \Pi_t\right)^2 }}\,\begin{pmatrix}
1\\
 \sqrt{\Pi_t^2 + 1} - \Pi_t
\end{pmatrix},
\end{equation}
where we introduced the notation $\Pi_t=(c \pi_t+A_t)/mc^2$. One easily convinces oneself that these eigenstates, $\Phi(p,\pi_t)$,  fulfill the orthonormality condition \eqref{eq13}.

For the following analysis we will assume that in the initial state, $\rho_0$, merely particles are present, but no antiparticles. This assumption is in full agreement with typical situations in nature and the mathematical treatment simplifies significantly \footnote{Antiparticles are characterized by their negative energies \eqref{eqa1}. Thus, the thermal distribution \eqref{eq24} would not be well-defined for positive temperatures. This conceptual issue appears to be sufficiently severe, that we decided to postpone the resolution of this problem to future work}. We emphasize that this assumption merely circumvents the conceptual issue of having to define a free energy for antiparticles. It has been shown that for any normalized initial state \cite{deffner_2013} a fluctuation theorem can be derived. However, such a general theorem only reduces to a generalized Jarzynski equality for thermodynamically well-defined situations \cite{deffner_2013}. Nevertheless,  for the sake of completeness, antiparticle energies and states can be found in Appendix~\ref{sec:appA}.

We also would like to emphasize that our analysis does not neglect the existence of antiparticles completely. We merely assume that the initial state is a thermal wave packet of particles. The dynamics, however, is described by the time-dependent Dirac equation \eqref{eq17}, and hence governed by both, positive and negative eigenenergies.  Hence, in particular the transition probabilities \eqref{eq04} are governed by both, particle and antiparticle solution.

\paragraph{Linear protocol}

As a first example we consider a linearly parameterized vector potential,
\begin{equation}
\label{eq21}
A_t=\alpha\, t\,,
\end{equation}
for which a solution of Eq.~\eqref{eq18} is given by
\begin{equation}
\label{eq22}
\begin{split}
\Psi_1(p,t)&=C_1(p)\,D_{-\nu}\left(\frac{(i+1)(cp+\alpha t)}{\sqrt{\alpha\hbar}}\right)\\
&+C_2(p)\,D_{\nu-1}\left(\frac{(i-1)(cp+\alpha t)}{\sqrt{\alpha\hbar}}\right)\,.
\end{split}
\end{equation}
Here, $D_\nu(\cdot)$ denotes the parabolic cylinder function \cite{abramowitz_1964} of order $\nu= i\,m^2 c^4/2\alpha\hbar$, and $C_1(p)$ and $C_2(p)$ are time-independent functions of momentum determined by the initial state. 

As mentioned earlier, for Dirac dynamics the solution is mathematically more involved, and also the amplitude of the wave function depends on the specific parameterization of $A_t$ -- not only the phase as in the previous, non-relativistic case \eqref{eq14}. This can be understood as \emph{dynamical interference} of the two components of the bispinor. Nevertheless, the transition probabilities can still be written as
\begin{equation}
\label{eq23}
\mf{p}_D\left(\pi_0\rightarrow\pi_\tau\right)=\left|\int\td p\,\Phi(p,\pi_\tau)\cdot\Psi(p,t)\right|^2\,\mf{p}_0^D(\pi_0)\,,
\end{equation}
where the initial distribution now reads
\begin{equation}
\label{eq24}
\mf{p}_0^D(\pi_0)=\e{-\beta\, \epsilon_D(\pi_0,A_0)}/Z^D_0\,,
\end{equation}
with which we can compute the quantum work distribution $\mc{P}_D(W)$ \eqref{eq03}.
\begin{figure}
\includegraphics[width=.48\textwidth]{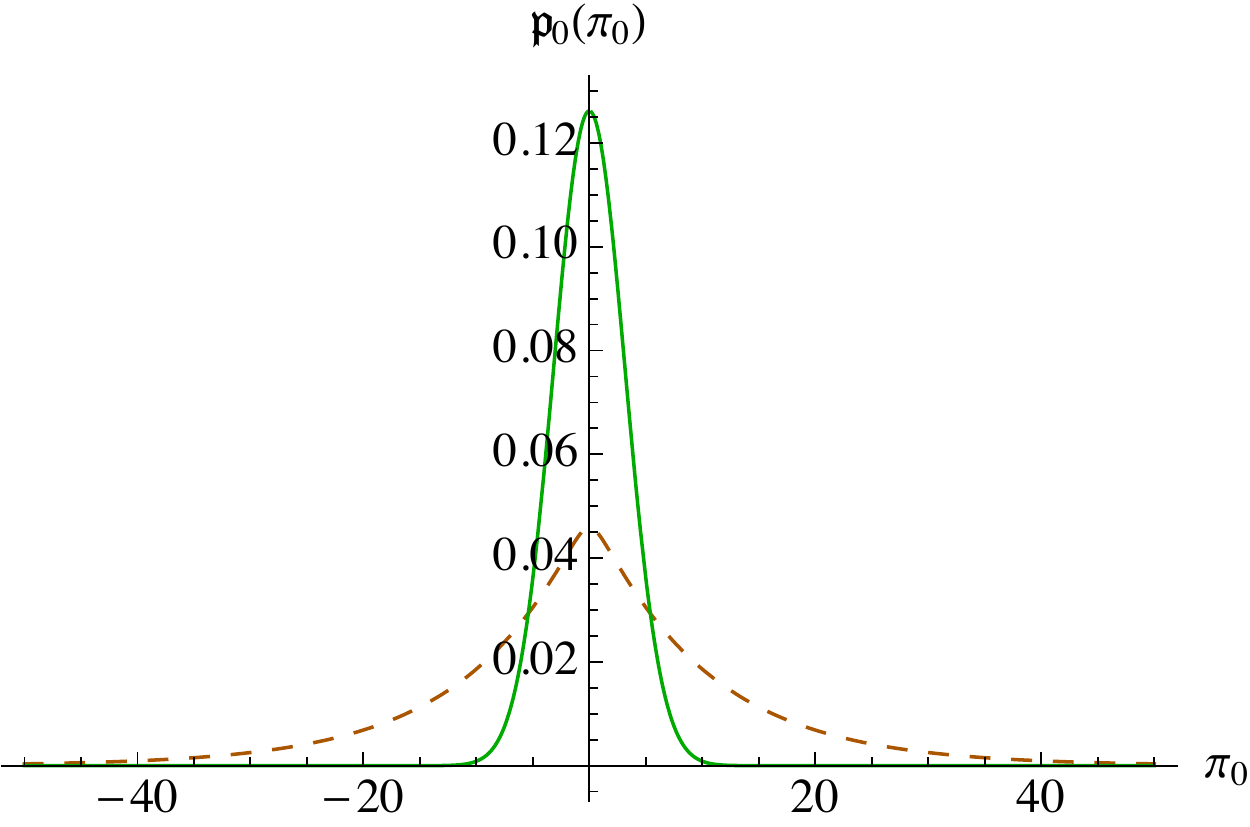}
\caption{\label{fig:thermal} (color online) Thermal momentum distribution for Schr\"odinger \eqref{eq15a} (green, solid line) and Dirac particles \eqref{eq24} (dark orange, dashed line) with $m=1$, $c=1$, $A_0=0$, and $\beta=0.1$.}
\end{figure}

In Fig.~\ref{fig:thermal} we plot the thermal momentum distribution for Schr\"odinger particles \eqref{eq15a} together with the distribution for Dirac particles \eqref{eq24}. In the Schr\"odinger case we have the well-known (Gaussian) Maxwell-Boltzmann distribution. The momentum distribution for relativistic Dirac particles is broader due to  the relativistic energy, $mc^2$, and was first described for classical mechanics by J\"uttner \cite{Juttner1911}. The so-called Maxwell-J\"uttner distribution converges towards the Maxwell-Boltzmann distribution \eqref{eq15a} for low temperatures, and decays slower than a Gaussian at high temperatures \cite{Juttner1911}. The major limitation is that the Maxwell-J\"uttner distribution neglects antiparticles, which however serves our present purpose.

\begin{figure}
\includegraphics[width=.48\textwidth]{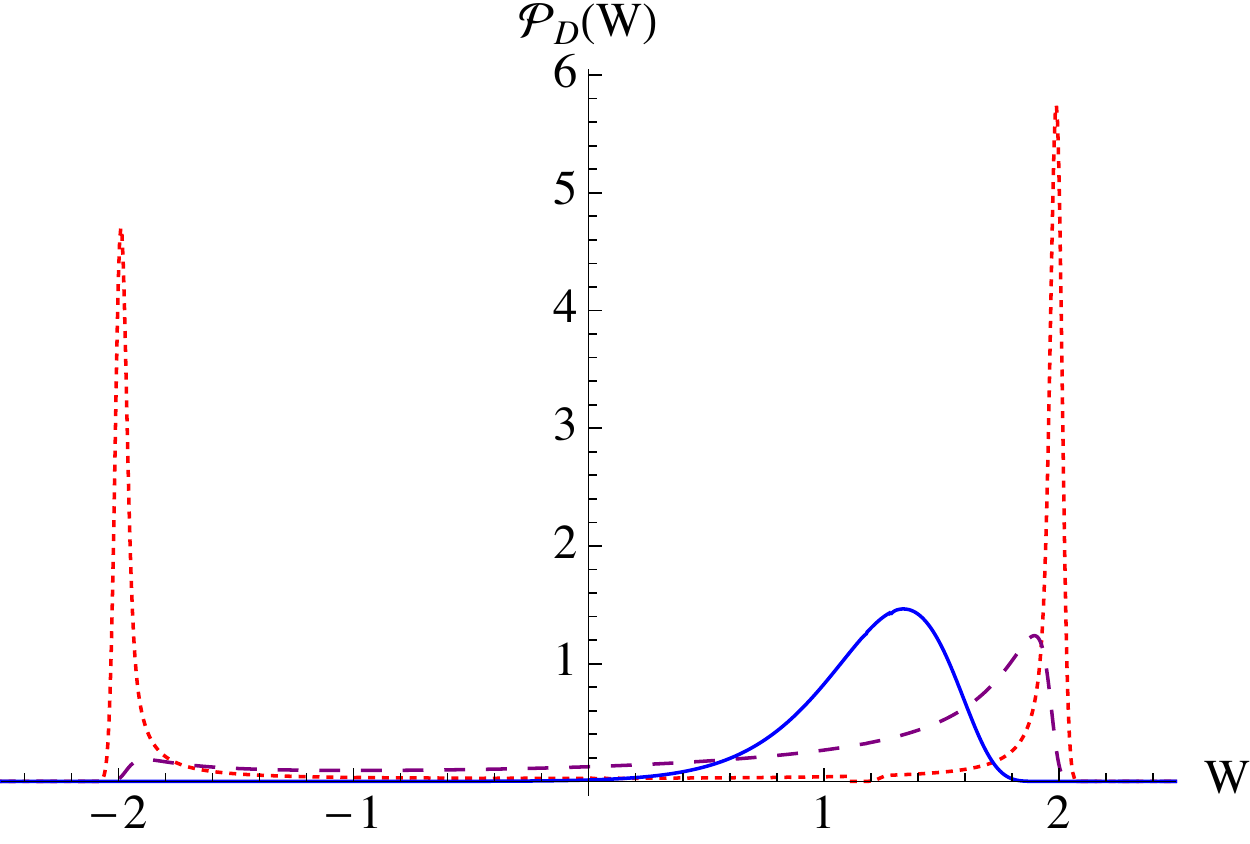}
\caption{\label{fig:dirac_lin} (color online) Quantum work density, $\mc{P}_D(W)$, for charged Dirac particles and the linear protocol \eqref{eq21} with $\alpha=1$,  $\tau=1$, $\hbar=1$, $m=1$, $c=1$, and $\beta=10$ (blue, solid line), $\beta=1$ (purple, dashed line), and $\beta=0.1$ (red, dotted line).}
\end{figure}

The resulting work distribution, $\mc{P}_D(W)$, is plotted in Fig.~\ref{fig:dirac_lin}, for the same parameters and color coding as for Schr\"odinger particles in Fig.~\ref{fig:schrodinger}. 
As anticipated the relativistic effects change the characteristics of the quantum work distribution significantly. The most striking difference with the Schr\"odinger case in Fig.~\ref{fig:schrodinger} is that $\mc{P}_D(W)$ has a finite support. This, however, can be understood intuitively: Large fluctuations in $W$ are accompanied by a large change in momentum. However, the momentum is limited by the light cone, and, hence, large fluctuations in $W$ are also ``cut-off'' by the light cone. 

 Quantitatively, the finite support can be determined by inspecting the transition probabilities \eqref{eq23}. It is easy to see that $\mf{p}_D\left(\pi_0\rightarrow\pi_\tau\right)\propto \de{\pi_0-\left(\pi_\tau+A_\tau/c\right)}$.  Hence, the only work values contributing to $\mc{P}_D(W)$ are $W=\epsilon_D(\pi_0+A_\tau/c,A_\tau)-\epsilon_D(\pi_0,A_0)$, and therefore $W\in (-2 \,\alpha \tau,\, 2\, \alpha \tau)$.

Qualitatively, one can understand Fig.~\ref{fig:dirac_lin} by starting with the distribution for Schr\"odinger particles in Fig.~\ref{fig:schrodinger}, and ``compressing'' the curves into the allowed support. For low temperatures (blue curve) the left flank is rather unaffected, as the distribution lives ``far away'' from the light cone, whereas the right flank is only slightly deformed. For higher temperatures the effect becomes more prominent, and the distribution becomes ``jammed'' at the edges of the support, i.e., at the light cone.

\paragraph{Exponential protocol}

To conclude the analysis we also compute the quantum work density \eqref{eq03} for a non-linear parameterization,
\begin{equation}
\label{eq26}
A_t=\alpha \left(1-\e{-t/\tau}\right)\,.
\end{equation}
Also in this case the time-dependent Dirac equation \eqref{eq18} can be solved analytically. However, the solution can no longer be written in compact form, and can be found in Appendix~\ref{sec:appB}. The transition probabilities \eqref{eq23} and the initial distribution remain the same by replacing Eq.~\eqref{eq22} with the expression \eqref{eqa4} everywhere.

\begin{figure}
\includegraphics[width=.48\textwidth]{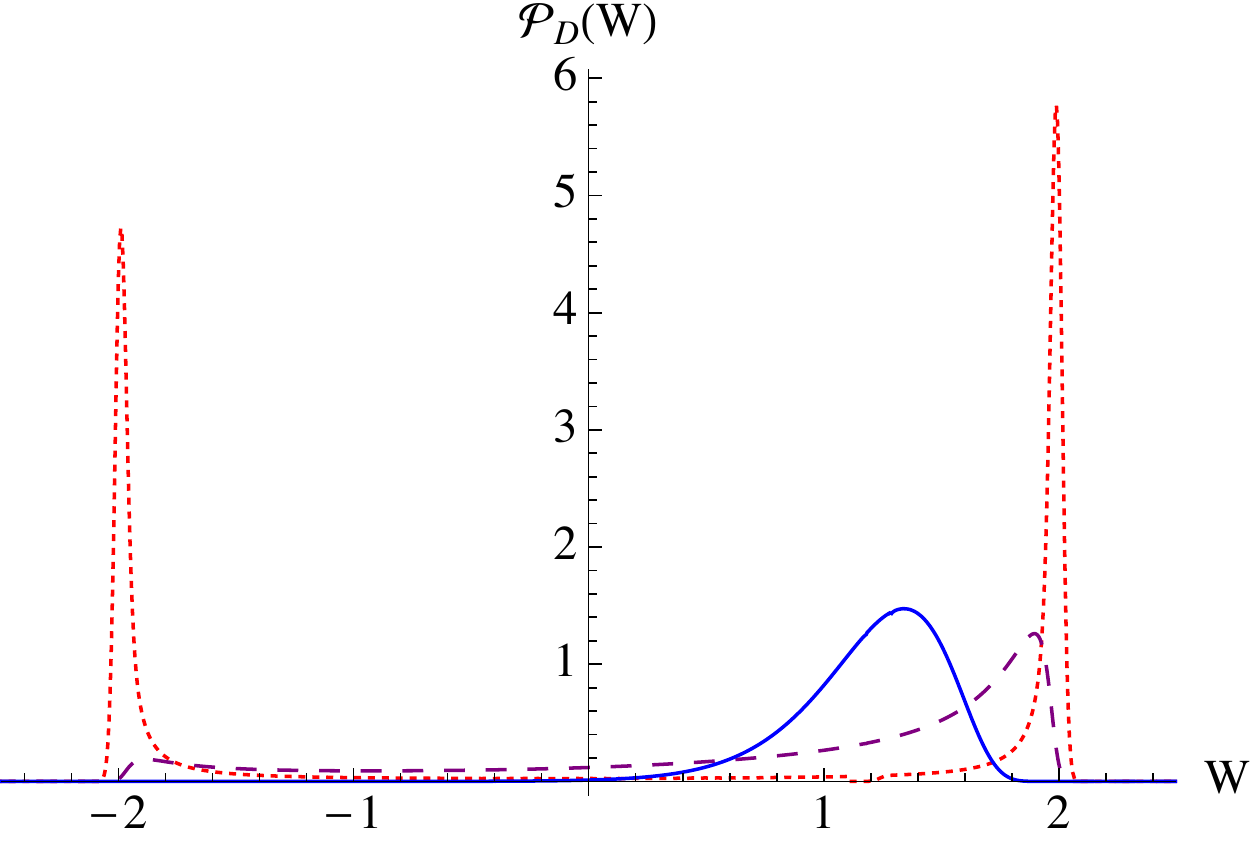}
\caption{\label{fig:dirac_exp} (color online) Quantum work density, $\mc{P}_D(W)$, for charged Dirac particles and the exponential protocol \eqref{eq26} with $\alpha=e/(e-1)$, $\tau=1$, $\hbar=1$, $m=1$, $c=1$, and $\beta=10$ (blue, solid line), $\beta=1$ (purple, dashed line), and $\beta=0.1$ (red, dotted line).}
\end{figure}

Figure~\ref{fig:dirac_exp} illustrates the resulting quantum work distributions. We observe that the work distributions resulting from the linear protocol \eqref{eq21} and the exponential protocol \eqref{eq26} are nearly indistinguishable -- despite the solutions \eqref{eq22} and \eqref{eqa4} being complicated expressions of special functions. Thus, we conclude that the effect of the light cone on the work distribution is more prominent than the interference of the two components of the bispinor \footnote{The indistinguishability of the work distributions for different protocols is a peculiarity of the very simple model. Thus it is reasonable to expect richer features for situations involving scalar potentials \cite{Fillion2013} or space-dependent vector fields.}.

\subsection{Jarzynski equality}

The validity of the quantum Jarzynski equality follows from the unitarity of Dirac dynamics. Nevertheless, it is worthwhile to numerically verify its predictions. To this end, we numerically integrated the average exponentiated work for the distributions in Figs.~\ref{fig:dirac_lin} and \ref{fig:dirac_exp}. Here, the Jarzynski equality becomes,
\begin{equation}
\label{eq27}
\la \e{-\beta W}\ra=\int\td W\,\mc{P}_D(W)\, \e{-\beta W}=1\,,
\end{equation}
as the free energy difference vanishes. In Table~\ref{tab1} we summarize the numerical results. We see that the quantum Jarzynski equality \eqref{eq27} is, indeed, verified to very high accuracy.
\begin{table}
\begin{tabular}{|c|c|c|c|}
\hline
&$\beta=10$ &$\beta=1$  &$\beta=0.1$ \\
\hhline{|=|=|=|=|}
linear &0.99 &0.99 & 0.99 \\
exponential &0.96  & 0.98 & 1.00\\
\hline
\end{tabular}
\caption{\label{tab1} Numerical verification of the quantum Jarzynski equality \eqref{eq27} for the quantum work distributions for Dirac particles in Figs.~\ref{fig:dirac_lin} and \ref{fig:dirac_exp}.}
\end{table}

The validity of the quantum Jarzynski equality \eqref{eq27} explains another important feature of $\mc{P}_D(W)$. For the linear \eqref{eq21} as well as for the exponential \eqref{eq26} protocol left and right flank of the distribution are ``exponentially asymmetric''. This asymmetry constitutes a necessary charactertistic of $\mc{P}_D(W)$ for Eq.~\eqref{eq27} to hold.  Note also that the asymmetry of $\mc{P}_D(W)$ is not an artifact of assuming that the initial state is comprised of only particles, but no antiparticles. We emphasize again that the existence of antiparticles is implicit in our treatment as the dynamics is described by Eq.~\eqref{eq17}.

\section{Concluding remarks}

In the present study we have analyzed the validity of the quantum Jarzynski equality and the properties of the quantum work distribution for systems described by the Dirac equation. For pedagogical reasons and for the sake of simplicity we focused on an illustrative case study. However, our system is more than a simple toy model, and it has realistic and experimental relevance.

\subsection{Experimental relevance}

Only recently, Fillion-Gourdeau \etal \cite{Fillion2012} studied the same model system in the context of pair production in counterpropagating laser light. However, Ref.~\cite{Fillion2012} not only solves Eq.~\eqref{eq17} analytically for an oscillating parameterization of $A_t$, but also provides relevant values for the field strength, for which the dynamics could be observed in an experiment.  It is worth emphasizing that Eq.~\eqref{eq09} is only an approximate description of the real physical situation with a clearly defined range of validity \cite{Fillion2012}. Nevertheless, for all experiments for which Eq.~\eqref{eq09} is valid our results could be readily verified, where one only would have to additionally measure the momentum distribution. From the momentum distribution one would compute the transition probabilities \eqref{eq23}, and build the quantum work distribution \eqref{eq03} from a histogram. This procedure is fully analogous to the cold ion trap  experiment, that verified the quantum Jarzynski equality  \cite{huber_2008,An2014b}. 

\subsection{Summary and outlook}

Our present analysis has extended the scope of quantum stochastic thermodynamics to relativistic energies. We have shown that not only does the quantum Jarzynski equality hold for Dirac dynamics, but we also have provided a step-by-step ``recipe'' of how to compute the relativistic work distribution.  For the sake of clarity and due to its mathematical simplicity we focused on free, charged particles traveling through a time-dependent vector potential. Another recent reference proposed to study pair production in a slightly more complicated, but also more realistic system including a scalar potential \cite{Fillion2013}. Our analysis could be straightforwardly applied to the situation of Ref.~\cite{Fillion2013} under the expense of having to compute $\mc{P}_D(W)$ fully numerically.

\acknowledgements{SD acknowledges financial support by the U.S. Department of Energy through a LANL Director's Funded Fellowship.}

\appendix

\section{\label{sec:appA} Antiparticle energy and eigenstate}

The instantaneous antiparticle solution of the time-dependent Dirac equation \eqref{eq17} is given by 
\begin{equation}
\label{eqa1}
\Phi^a(p,\pi_t)=\frac{\de{p-(\pi_t+A_t/c)}}{ \sqrt{1 + \left(\sqrt{\Pi_t^2 + 1} + \Pi_t\right)^2 }}\,\begin{pmatrix}
-1\\
 \sqrt{\Pi_t^2 + 1} + \Pi_t
\end{pmatrix}.
\end{equation}
with eigenenergies
\begin{equation}
\label{eqa2}
\epsilon_D^a(\pi_t, A_t)=-\sqrt{\left(c \pi_t+A_t\right)^2+\left(m c^2\right)^2}\,.
\end{equation}
Note that the eigenenergies for antiparticles are negative. Thus, the Maxwell-J\"uttner distribution \eqref{eq24} is ill-defined for positive temperatures.

\begin{widetext}
\section{\label{sec:appB}Analytical solution of time-dependent Dirac equation}

For the exponential protocol,
\begin{equation}
\label{eqa3}
A_t=\alpha \left(1-\e{-t/\tau}\right)
\end{equation}
a solution of the time-dependent Dirac equation \eqref{eq18} is given by
\begin{equation}
\label{eqa4}
\begin{split}
&\Psi_1(p,t)=\e{-\frac{i}{\hbar}\left(t\,\sqrt{(m c^2)^2+(c p)^2+2 pc \alpha+\alpha^2}+\alpha \tau\, e^{-t/\tau}\right)}\\
&\times\bigg[C_1(p)\,U\left(-\frac{i \tau}{\hbar}\left(cp+\alpha-\tau\sqrt{(m c^2)^2+(cp)^2+2cp\alpha+\alpha^2}\right),\,1+\frac{2i\,\tau}{\hbar}\sqrt{(m c^2)^2+(cp)^2+2cp\alpha+\alpha^2},\,\frac{2 i\,\alpha\tau}{\hbar}e^{-t/\tau}\right)\\
&+C_2(p)\, L\left(\frac{i \tau}{\hbar}\left(cp+\alpha-\tau\sqrt{(m c^2)^2+(cp)^2+2cp\alpha+\alpha^2}\right), \frac{2i\,\tau}{\hbar}\sqrt{(m c^2)^2+(cp)^2+2cp\alpha+\alpha^2},\,\frac{2 i\,\alpha\tau}{\hbar}e^{-t/\tau}\right)\bigg]\,.
\end{split}
\end{equation}
Here, $U(\cdot,\cdot,\cdot)$ is the Kummer function, and $L(\cdot,\cdot,\cdot)$ denotes the Laguerre polynomial \cite{abramowitz_1964}.
\end{widetext}

\bibliography{jar_dir_bib}

\end{document}